\documentclass[conference, a4size, 10pt]{IEEEtran}
\IEEEoverridecommandlockouts
\usepackage{cite}
\usepackage{amsmath,amssymb,amsfonts}
\usepackage{algorithmic}
\usepackage{graphicx}
\usepackage{textcomp}
\usepackage{xcolor}
\def\BibTeX{{\rm B\kern-.05em{\sc i\kern-.025em b}\kern-.08em T\kern-.1667em\lower.7ex\hbox{E}\kern-.125emX}}
    
\usepackage{acronym}
\usepackage{siunitx}
\usepackage{cuted}
\usepackage{subfigure,epsfig,xcolor, pstricks}
\usepackage{todonotes}
\usepackage{balance}

\DeclareSymbolFont{EULE}{U}{eur}{m}{n}
\DeclareMathAlphabet{\matheule}{U}{eur}{m}{n}
\DeclareMathSymbol{\dirac}{\mathalpha}{EULE}{"0E}

\begin{document}

\title{Towards Realistic Statistical Channel Models For Positioning: Evaluating the Impact of Early Clusters}

\author{
    Mohammad Alawieh\IEEEauthorrefmark{1},
    George Yammine\IEEEauthorrefmark{1}, 
    Ernst Eberlein\IEEEauthorrefmark{1}, 
    Birendra Ghimire\IEEEauthorrefmark{1}, 
    Norbert Franke\IEEEauthorrefmark{1}, 
    Stephan Jäckel\IEEEauthorrefmark{2},\\
    Tobias Feigl\IEEEauthorrefmark{1},
    and Christopher Mutschler\IEEEauthorrefmark{1}\\
    \and
    \IEEEauthorblockA{
        \IEEEauthorrefmark{1}~Fraunhofer Institute for Integrated Circuits IIS, N\"urnberg, Germany\\
        \texttt{\footnotesize\{firstname.lastname\}@iis.fraunhofer.de}
    }
    \and
    \IEEEauthorblockA{
        \IEEEauthorrefmark{2}~SJC Wireless, Berlin, Germany\\
    	\texttt{\footnotesize jaeckel@sjc-wireless.com}
    }
}

%% Story line (proposal ebl)
%  
%% === characteristics of BW limited correlation 
%  - Existing channel models focus on communication applications
%  - According to the Nyquist theorem the channel can be represented with a sampling rate according to the bandwidth
%  - If the bandwidth limitation (and no further pulse shaping filter) is applied for a ideal channel (with dirac impulse as CIR) the measured CIR is equivalent to a SINC function. 
% ==> Fig 1
%  - if the true not bandwidth limited CIR includes several DIRAC the CIR an be represented by the complex value weighted sum of several SINC function 
% ==> fig 2: Example for sum of 2 SINC functions, the second function is deleyed by 3ns (0.9m) and attenuated by 6dB (magnitude 0.5). The result depends on the phase. Figure shows two examples: inphase (same phase as LOS) and opposite phase. 

%%
%  - For positioning purpose super resolution techniques try to estimate the ToA of the first arriving path (FAP) with a resolution better than the sampling rate taking. 
% - To into account that no signal arrive before the delay according to the distance. 

\maketitle

%%%%%%%%%%%%%%%%%% abstract %%%%%%%%%%%%%%%%%%%%%%%
\begin{abstract}
Physical effects such as reflection or diffraction cause a radio signal to travel from a transmitter to a receiver in multiple replicas of different amplitude and rotation. Early clusters define replicas from objects in the immediate vicinity of a transmitter or a receiver, which influence the lobe of the  Line-Of-Sight (LOS). For positioning in particular, the replica, which arrives with a delay bordering that of the LOS path highly impacts the performance evaluation and must not be overlooked in the channel model considerations.

In this paper, we show that the existing channel model specified for performance evaluation within 3GPP does not cover the above phenomena with sufficient accuracy and simulation results deviate significantly from the measured values. As an extension to overcome this shortcoming, we propose model early clusters to simulate measured effects with sufficient accuracy and ensure that it channel model reflects typical effects  in real scenarios. 

\end{abstract}

%%%%%%%%%%%%%%%%%% keywords %%%%%%%%%%%%%%%%%%%%%%%

%%%%%%%%%%%%%%%%%% intro %%%%%%%%%%%%%%%%%%%%%%%
\section{Introduction}
\label{sec:Introduction}
% Status
% - First draft written by ebl
% - Reference to hybrid model as described in TR38.901
% General 3GPP positioning evaluations with reference to TR38.857 and TR38.855 and how its progressing to include precise positioning and AI aspects (if we can talk down on cluster correlation from multiple BSs, we highlight this as an enabler for FP (fingerprint) evaluation approaches) . 

\begin{figure}[!tp]
    \centering
    \includegraphics[trim=25 2 25 2, clip, width=0.9\columnwidth] {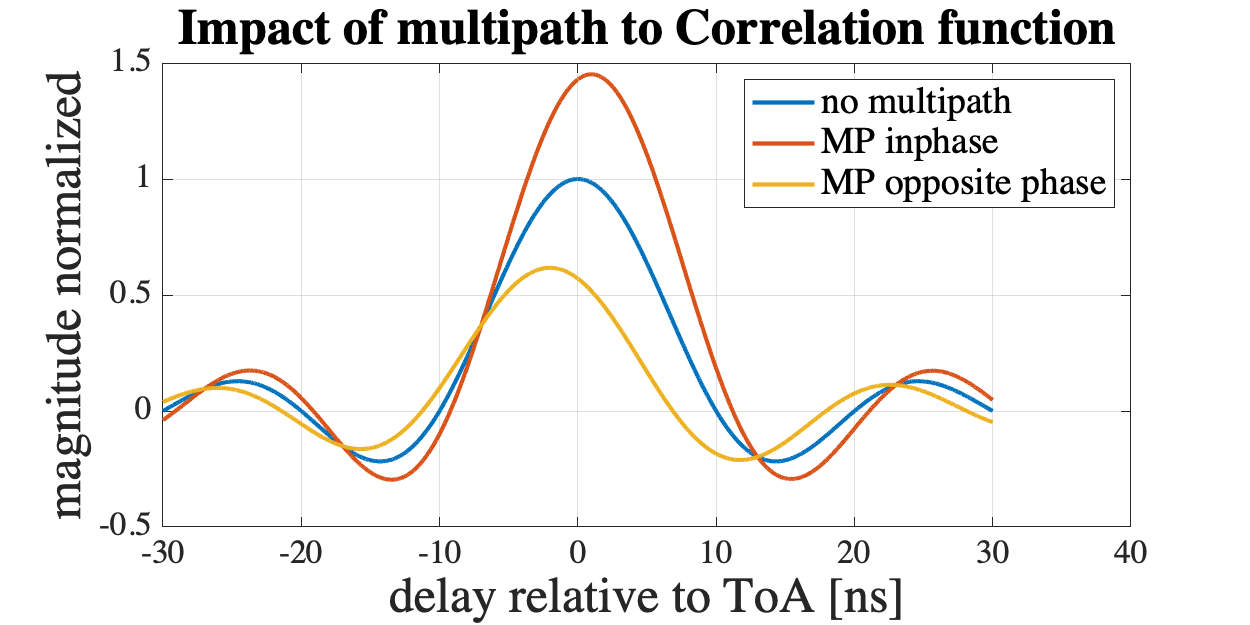}
    \caption{The first lobe of the correlation function. 
    The correlation peak deviates from its true value when multipath signals superimpose.}
    \vspace*{-0.35cm}
    \label{fig:corrMagEC}
\end{figure}

\noindent
Signals traversing a multipath channel consist of multiple replicas of the original signal that have been attenuated and phase rotated.
The channel impulse response (CIR) can be estimated by correlating the received signal with the transmitted sequence. In the case of signals with limited bandwidth, the correlation paths related to the LOS and different multipath components (MPC) may overlap (constructively or destructively) and it is unlikely to separate them.
%Enhanced algorithms are required to achieve time resolution that is more accurate than resolution that is inversely proportional to signal bandwidth.
%Thus, the MPCs that arrive within the temporal resolution of the signals cannot be deciphered from one another and are either constructively or destructively superimposed at the receiver.
As a result, the correlation peak deviates from its true value, see Fig.~\ref{fig:corrMagEC}.
The deviation corresponds to the geometric distance of the LOS between the transmitter and the receiver and introduces errors in time-of-arrival (ToA) estimates, and thus degrades position accuracy.
%For LTE, a positioning error below 50\,\si{\m} is defined for regulatory use cases and below 3\,\si{\m} for commercial use cases.
Depending on the signal bandwidth and the target accuracy requirements, it may be not necessary to consider paths arriving with similar delay. But for high accuracy ToA estimation or for limited bandwidth, realiable ToA estimation algorithms must be able to detect the ToA of the first arriving path even if the related correlation peak is impaired by other multipath components. Moreover, 3GPP has gradually increased the requirements of the 3GPP-NR standard w.r.t. accuracy, latency and reliability.
In particular, the support of service classes 5 or 6~\cite{tr22104,tr2261} require an accuracy of less than 30\,\si{cm} (ToA error of 1\,\si{ns}).
Reflections from objects close to the receiver or transmitter may have a delay of a few \si{ns} or even lower. Typical examples are the ground reflection or object close to the mounting point of the antennas. 

%The evaluation of a positioning algorithm yields plausible results when essential aspects of the deployment environment are modeled correctly, while non-essential effects are abstracted to save computational resources.
To evaluate positioning performance, ~\cite{tr38857} and ~\cite{tr38885}  specified several scenarios using geometry based statistical channel models. These channel models are based on measurements and focused on effects relevant for data communication applications. Effects relevant for positioning may be not well covered or the models offer limited flexibility to adjust the related parameters.
For communication applications typically CIRs with high delay spread are more critical, whereas the for positioning especially the characteristics of the first arriving paths are essential. 

To overcome these issues, we propose methods allowing a better control of the properties of each link.  
Additional enhancements supporting a better evaluation of machine-learning based algorithms and emulation of "obstructed LOS" (OLOS) conditions are in the scope of \cite{SDCpaper}.  Accordingly, we consider early (arriving) clusters (EC) and late arriving clusters (LC). We developed a modified TR38.901 model offering a more flexible adjustment of the ratio between the EC and LC. It is the nature of statistical models that EC are generated also randomly. However, to better control the probability and the strength of the EC additional parameters are required. Furthermore, we propose a metric allowing the evaluation of the performance versus channel characteristics. This helps to identify critical channel conditions.

%Furthermore, in this scenario, infrastructure elements, such as transmission and reception points, are deployed such that coverage is enhanced. 
%However, the deployment may not be optimal for computing UE position.

%Some of the issues have been addressed by a hybrid model \cite{tr38901}, which combines ray-tracing and statistical model. 
%In this model, the scatting clusters generated by the ray-tracing model are merged with clusters generated randomly by the statistical model. 
%However, the hybrid approach \cite{tr38901} is typically not used as basis of system evaluation because of the complexity of modelling and the lack of availability of additional details of environment (e.g., 3D maps). 

%The model can be simplified to capture the propagation effects that are relevant to evaluate the positioning algorithms. 

The rest of this paper is structured as follows. 
Sec.~\ref{sec:Stateofart} gives a short introduction to the principle of the TR38.901 model. 
Sec.~\ref{sec:extensionstochannelmodels} describes our proposed extension  of the existing channel model Sec.~\ref{sec:InfluenceOfEC} evaluates the effects caused by ECs.
Sec.~\ref{sec:Measurements} compares measured CIRs with CIRs generated in simulation with and without the proposed extensions.

\section{State of the Art}
\label{sec:Stateofart}
% \todoALLIn{motivate with 3GPP model and PDP.}

\noindent
Most research in the field of communications engineering relies on modeling the wireless radio propagation and environment. For comparison of the results of studies performed by different companies, 3GPP defines a standardized channel model in Technical Report TR38.901~\cite{tr38901} including deployment scenarios, channel parameters and characteristics of key components as antennas, for example. 
The underlying statistical model shall generate propagation conditions representing different usage scenarios. The generated channels shall cover ``easy'' scenarios as well as critical scenarios with a probability as expected in the given deployment. 
For positioning algorithms using several base stations and triangulation, for example, the performance may already degrade, if one link provides a low accuracy. For the evaluation of the reliability of a technology very long simulations may be required, ensuring that critical scenarios are well covered, also.   

The TR38.901 model splits the CIR in clusters, where each cluster being composed of several subpaths. The models define the statistical properties of the cluster delays, AoA, AoD, etc. and the CIRs are generated randomly according these parameters. 

To simulate a specific radio system setup, the geometric layout (position of base stations, etc.) and the environment parameter set (called scenario table) are selected. The scenario tables cover the statistical properties of the channel characteristics defined by parameters like delay spread, angular spread, the Ricean K-factor, and shadow fading.
The CIRs itself are generated in a two step approach: 
\begin{itemize}
\item The large scale model emulates the changes of the propagation conditions and generates the instantaneous values of the delay spread, K-factor, shadow fading, etc. inline with the statistical distributions defined by the scenario tables. 
\item The small scale model generates the CIRs itself, controlled by the values generated by the large scale model. 
\end{itemize}
The channel model generates the multipath components of the power-delay-profile (PDP) of the CIR composed of $N$ clusters by randomly selecting for each cluster $n$ the delay $\tau_{\mathsf{n}}$ by
\begin{equation} \label{eq:DELAY}
    \tau_{\mathsf{n}} = -r_{\mathsf{\tau}} DS \ln\left( X_\mathsf{n} \right) \;,
\end{equation} 
and the cluster power $P_{\mathsf{n}}$ by
\begin{equation} \label{eq:PDP}
    P_{\mathsf{n}} = \exp \left( -\tau_{\mathsf{n}} \frac{r_\mathsf{\tau}-1} {r_\mathsf{\tau}DS} \right) 10^\frac{Z_\mathsf{n}} {10} \;,
\end{equation} 
where $DS$ is the spatially-correlated delay spread generated by the large scale model, $X_\mathsf{n}$ is a uniformly-distributed random variable and $Z_\mathsf{n}$ is a normal-distributed random variable implementing the per cluster shadow term. $r_\mathsf{\tau}$ is a proportionality factor defined by the scenario tables. These equations offer limited flexibility to generate PDP covering effects like reflections from objects close to the transmitter or receiver or other effects relevant for high accuracy ToA estimation. For the evaluation of positioning algorithms, it is essential to pay attention to the characteristics of the components arriving with a low delay---called early (arriving) cluster (EC). 
With a certain probability the equations \eqref{eq:DELAY} and \eqref{eq:PDP} generate also EC. But the probability depends on the number of clusters $N$ and the $DS$. This means, it is not possible to adjust the parameters related to effects resulting from reflections close to the transmitter and receiver and the overall delay spread independently.

\section{Extension of the TR38.901 Model}
\label{sec:extensionstochannelmodels}
\begin{figure}
    \includegraphics[width=1.0\columnwidth] {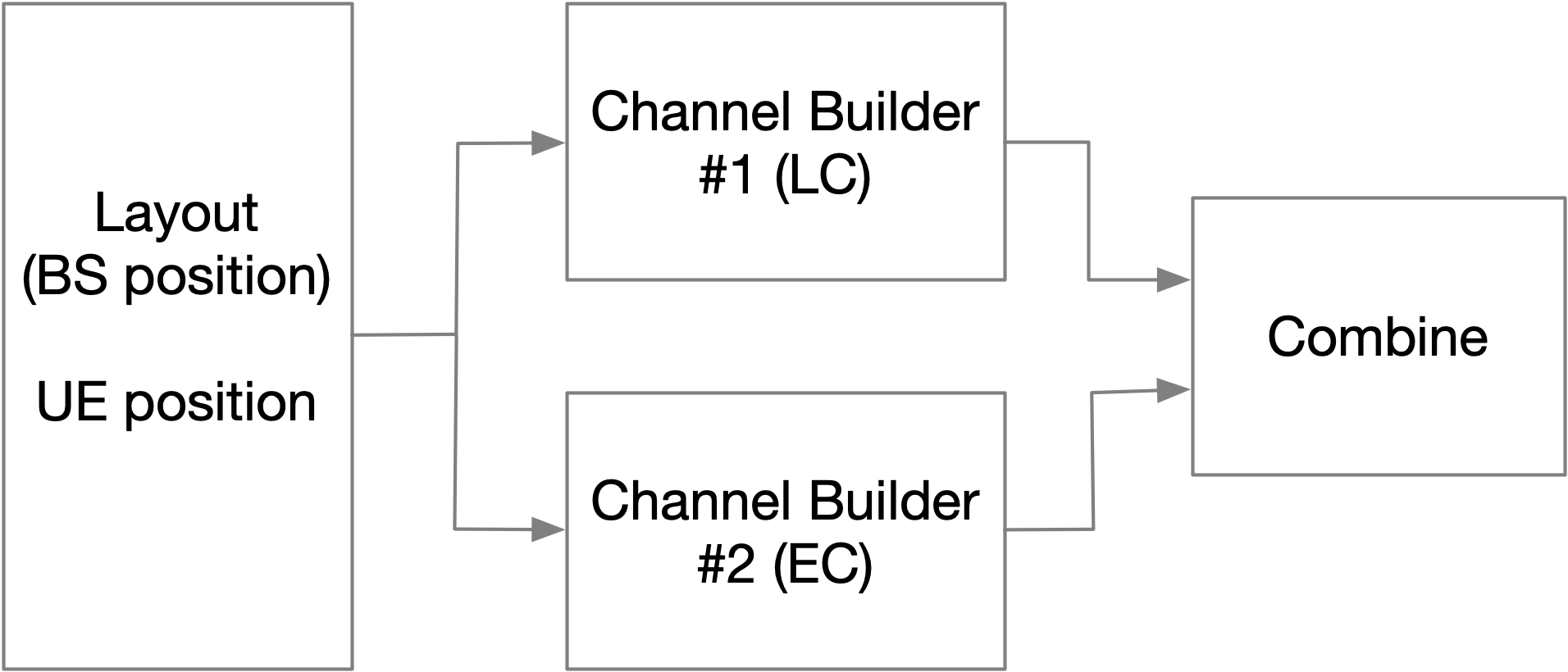}
    \caption{Extension of the TR38.901 model using two channel builders.}
    \label{fig:TwoBuilder}
\end{figure}

\noindent
Fig.~\ref{fig:TwoBuilder} depicts the principle of the proposed extension of the TR38.901 model. The CIR is generated by combining two CIRs generated by two channel builders as defined by TR38.901. Two sets of clusters are generated using parameter sets focusing on the components with high path delay (LC) and clusters arriving early (EC). The combining includes a normalisation to maintain the overall power according to the pathloss and shadow fading and (optional) other properties like the overall K-factor. The parameter sets for each builder can be set independently. The ratio between the LC and EC can be selected. 
To demonstrate the effect we selected four configurations (marked with configuration numbers)
\begin{itemize}
    \item 104.20 represents the parameter set ``InF LOS'' as defined by TR38.901
    \item 104.60 adds the ground reflection according the method as already supported by TR38.901
    \item 104.63 uses two builders. The combiner is set to ``maintain overall K-Factor statistics''
    \item 104.66 adds to the 104.63 configuration the ground reflection. 
\end{itemize}

\begin{figure}
    \includegraphics[width=1.0\columnwidth] {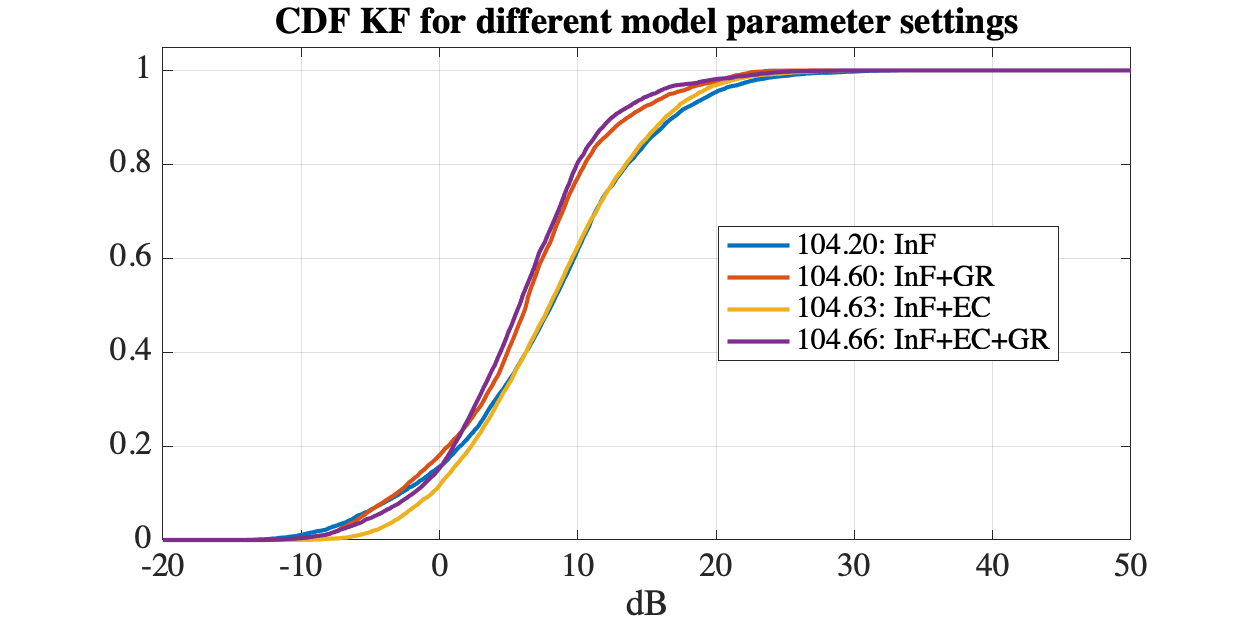}
    \caption{CDF of the overall K-factor for different model parameters. 
    The statistics are maintained when increasing the $K_{\mathsf{EC}}$ probability.}
    \label{fig:CDF_KF}
\end{figure}

\noindent
From the K-factor statistic given in Fig.~\ref{fig:CDF_KF} we observe that 104.20 and 104.63 have the same K-factor properties. In addition, ground reflection according TR38.901 may influence the overall K-factor. Along with CIRs with low (less than 0\,\si{dB}) K-factor are generated with a probability of circa 10 percent.

\noindent
To further evaluate the channel characteristics with respect to the EC we selected a modified K-factor (which we denote as $K_{\mathsf{EC}}$) as criterion. The $K_{\mathsf{EC},\Upsilon}$ considers the signal components arriving within a time window $\Upsilon$ relative to the line-of-sight path. 
This is given as 
\begin{equation} \label{eq:KFEC}
    K_{\mathsf{EC},\Upsilon} = 10 \log_{10} \left( \frac{P_\mathsf{LOS}}{\sum_n P_\mathsf{MP,n}} \right) \;, \quad n \in \mathcal{N} \;,
\end{equation} 
where $\mathcal{N}$ is the set of indices of clusters having a delay less than $\Upsilon$, i.e., $\tau_n < \Upsilon$, 
$P_\mathsf{LOS}$ denotes the power of the LOS component and $P_\mathsf{MP,n}$ denotes the power of the $n$th NLOS cluster. 
The cluster powers are as defined in the 38.901 model \cite{tr38901}. 

\begin{figure}
    \includegraphics[width=1.0\columnwidth] {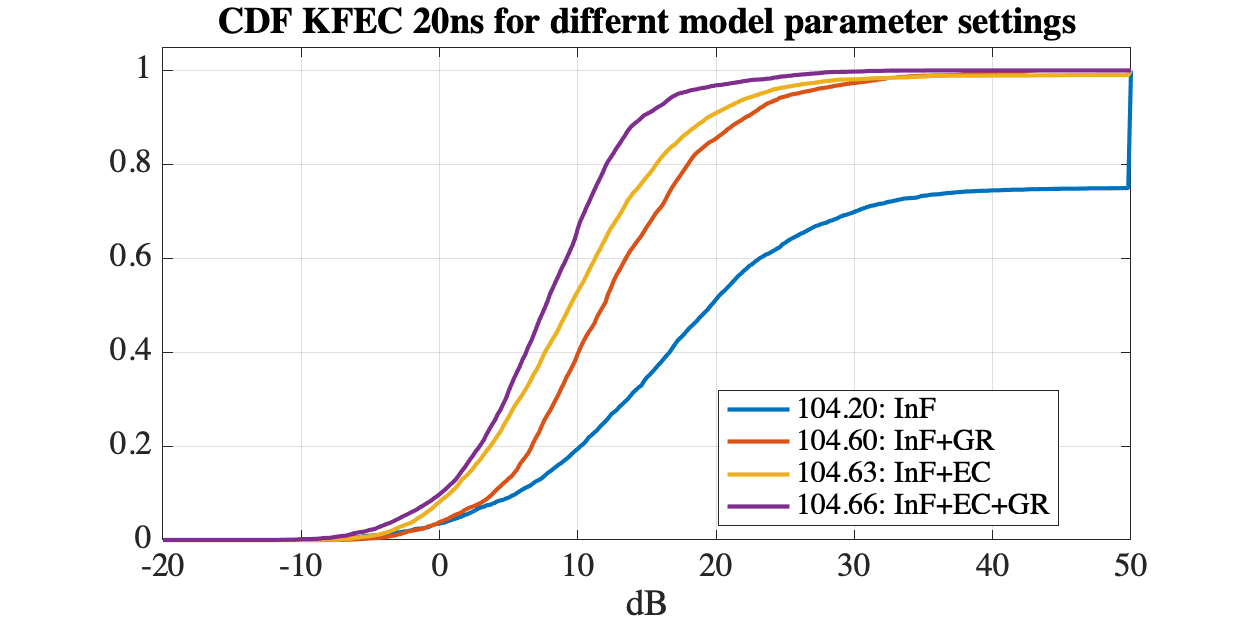}
    \caption{CDF of the $K_{\mathsf{EC}}$ for different model parameters. 
    The basis scenario (104.20) is the InF-LOS from 38.901. 
    The effect of ground reflections (GR) and early clusters (EC) on the measured $K_{\mathsf{EC}}$ is clearly seen.}
    \label{fig:CDF_KFEC20ns}
\end{figure}

In Fig.~\ref{fig:CDF_KFEC20ns}, we depict the CDFs of the $K_{\mathsf{EC}}$ statistics for different channel model parameter settings as given above. For InF LOS, the $K_{\mathsf{EC}}$ in 25 percent of the cases no multipath component arrives within a delay of 20ns. This means multipath components with a delay typical for ground reflection and in a radius in the range of 5m no objects reflecting the signal are placed. 

Fig.~\ref{fig:CDF_KFEC20ns} also show that using EC allows to change the $K_{\mathsf{EC}}$ while maintaining the overall K-factor.

% figure generated with "ML_evalHDF5_paper.m  , set 7.1"

% figure generated with "ML_evalHDF5_paper.m  , set 7.1"

%%%%%%%%%%%%%%%%%% problem description %%%%%%%%%%%%%%%%%%%%%%%
\section{Influence of Early Clusters on ToA Estimation Accuracy}
\label{sec:InfluenceOfEC}
%%  Section ChannelCharacteristics
%
% Link-level (one link (= one TRP with one or more antennas) and one UE at random position) scenarios
% - "OLOS scenarios"
%    - Objects in the fresnel zone (examples  ==> photos of 
%      experimental setups etc. ??? Relcovair photos )
%    	- Object of the size  ... 1m ... 2m .... are considered   
%         either as "blocker" if close to the UE or TRP
%         or as object with "diffraction" etc.
%        - Examples are industrial scenarios (ATVs, racks, hanging 
%          stage, forklift....)  or cars (e.g. platooning scenario) 
%    - Typical numbers for additonal path delay  ==>  "early cluster" %      definition
%      ==> first delayed path arrives within  n*1/bandwidth   
%          (n = 1...3) or fixed values  (10ns ... 20ns)
%- Other "early cluster"
%    - Ground reflections 
%    - "waveguide-scenarios" (street canyon)
%
% System level scenarios (several TRP, correlation of the 
% "reception state" ==> AoA dependent blockage probability
% - For 38.901 no AoA dependent LOS/NLOS modelling 
% - LOS/NLOS modelling ==> motivates Ray-Tracing
%
% - Probability of early cluster 
%    - compare values with examples given above (e.g. for InF: in 50 percent of the cases no ground reflection)
% - OLOS not supported ==> introduce first possible methods to generate OLOS as "LOS with low K-factor" or NLOS with low ATOA delay) 

%- trade-off "cluster density" (and probability of early cluster) versus delay spread and Num-Clusters
% - Limited flexiblity to define PDP (power delay profile)

\noindent
To demonstrate the impact of the EC, we ran simulations of a ToA estimator optimized for detecting the first arriving path of a 5G reference signal. The algorithm is used as an example for a super resolution algorithm and is based on detecting the inflection point of the rising edge of the correlation function which outperforms peak-detection-based methods.  
Here, SRS according the 5G NR standard was used with a signal bandwidth set to 100\,MHz and sampled at 122.88\,MHz at the receiver.
For each randomly generated channel realization, first, we calculated $K_{\mathsf{EC}}$, cf.\ \eqref{eq:KFEC}. For this particular example, we set $\Upsilon = 20\,\mathrm{ns}$. 
To cover a wide range of $K_{\mathsf{EC}}$) we combined the results of several simulation runs using the four channel model parameter settings as given above. We sorted the ToA estimation errors according the $K_{\mathsf{EC}}$ criterion and generated for different $K_{\mathsf{EC}}$ ranges separate ToA error CDFs

\begin{figure}
    \includegraphics[width=1.0\columnwidth] {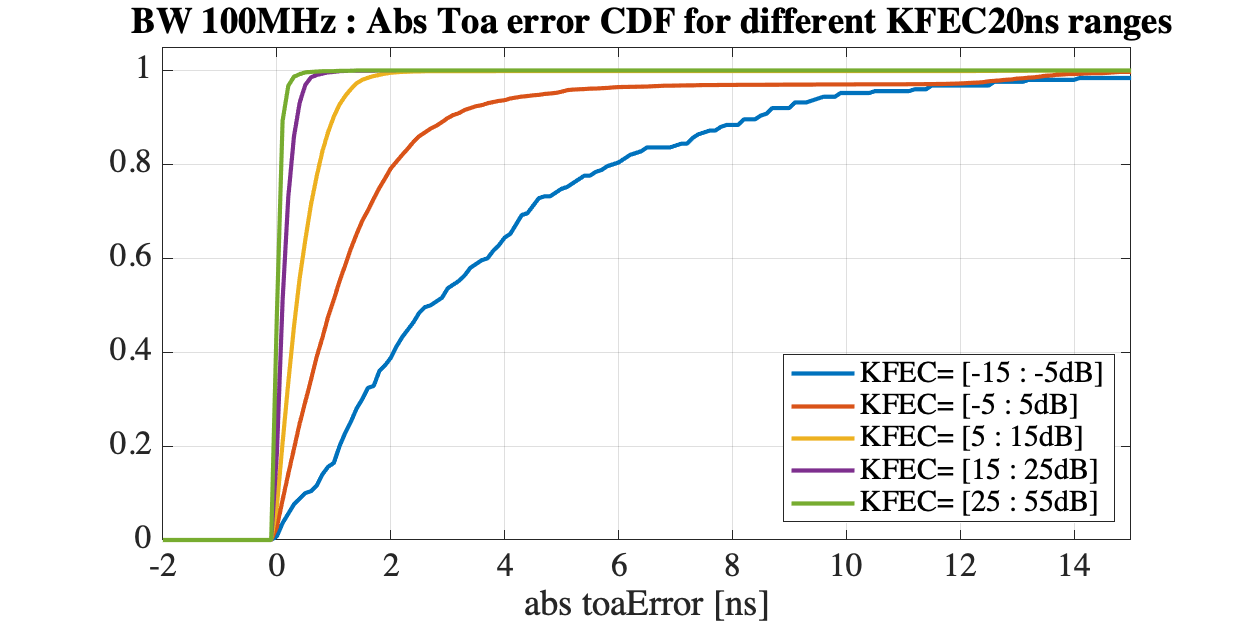}
    \caption{Empericial CDFs of the ToA estimation error for different groups of $K_{\mathsf{EC}}$ ranges using a 5G reference signal with a bandwidth of 100\,MHz. 
    In this example, $\Upsilon = 20\,\mathrm{ns}$ is used.}
    \label{fig:ToAerrorCDFsorted}
\end{figure}

The obtained empirical CDFs are all plotted in Fig.~\ref{fig:ToAerrorCDFsorted}. 
Here, we see that $K_{\mathsf{EC}}$ has a high impact on the ToA estimation accuracy. 
In fact, we have a large variation in the observed error. 
When $K_{\mathsf{EC}}$ is high, any reflecting object must be placed at a distance of at least 3 meters (when behind the transmitter or receiver) to ensure that all multipath component arrived with a delay higher than $20\,\mathrm{ns}$.

It should be noted that the grouping of the results according the calculated $K_{\mathsf{EC}}$ offers two advantages: 
\begin{itemize}
    \item It is possible to identify the scenarios where a higher error can be expected or the method fails. 
    \item The simulation results are less dependent on the overall channel statistics. The overall channel statistics has mainly an impact to the share of channel instances per $K_{\mathsf{EC}}$ group. 
\end{itemize}

%%%%%%%%%%%%%%%%%% Measurement Setup %%%%%%%%%%%%%%%%%%%%%%%
\section{Measurement and Evaluation}
\label{sec:Measurements}
% - Evaluation of RT Delay statistics and CIR
%
% % ==> Observations: 
% - new parameter sets are required
% - more flexibility for PDP profile definition 
% - pure statistical model:  Issues to match statistics with characteristics "surrounding area"
%
% ==> introduce SDC concept
% - Three methods for SDC position definition 
%    -- statistical model 
%    -- definition according usage scenario 
%    -- combination of both
%

\begin{figure}[b]
    \centering
    \includegraphics[width=1.0\columnwidth] {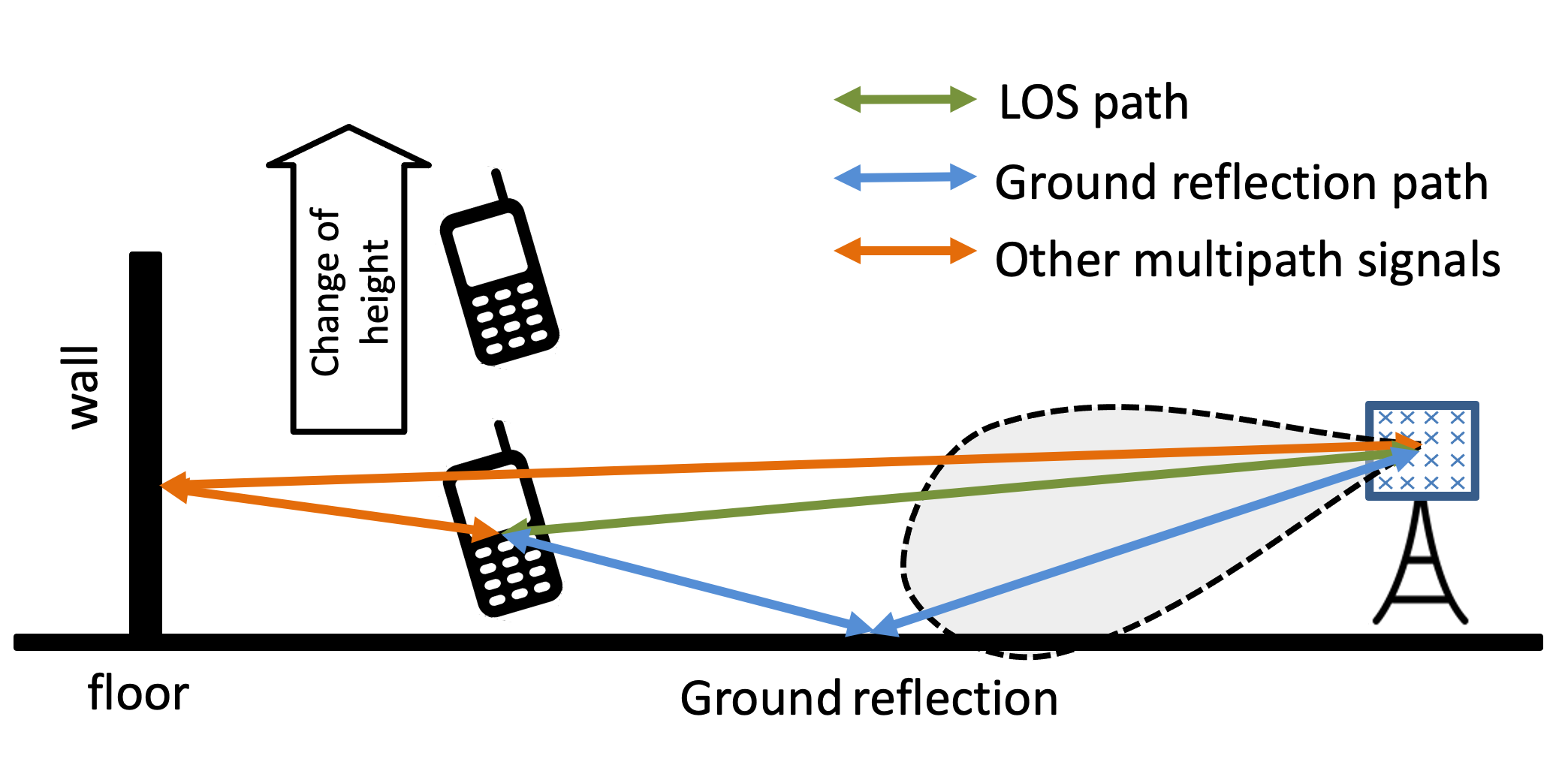}
    \caption{Ground reflection experiment setup. 
    The BS location is fixed and the height set at 1.7\,m. 
    The UE's height is varied between 0.7\,m and 3.3\,m.}
    \label{fig:GRexperimentSetup}
\end{figure}  

\noindent
We conducted a measurement campaign to demonstrate the behavior of the TR38.901 model with and without extension and to validate the impact of early clusters.
The comparison of measurements with simulations using statistical models can be only performed if the measurements covers also significant amount of different environment characteristics. This is time consuming and subject of future work. For the comparison of the CIRs generated by simulation with an experimental setup the channel model statistics must be constraint accordingly to achieve a better match with the experimental setup. The experimental setup can be considered as small subset of real world scenarios, whereas the statistical model targets several scenarios. In our experiment we focused on the impact of one type of EC, the ground reflection.

The measurement setup is described in  Fig.~\ref{fig:GRexperimentSetup}, and the campaign  was carried out in the L.I.N.K.\ test center \cite{linkhall}. 
The general planning of the floor consists of an open area bounded by the walls. In a first experiment we placed no additional objects around the UE. Hence, we expect mainly EC from ground reflections (GR) and ceiling. The distance to the walls was around 10\,\si{m}. These reflections are not considered as EC. 
The measurements were performed at a carrier frequency of 3.75\,\si{GHz} (wavelength = 8\,\si{cm}). To change the additional path length of the ground reflection we moved the UE along the Z-axis covering a UE height in the range between 0.7\,\si{m} and 3.3\,\si{m}. The different UE height has a minor impact to the LOS distance and the beamwidth of the antenna was high to minimize antenna effects on the pathloss. The BS was positioned at a height of 1.7\,\si{m}.
The measured channels that were analyzed correspond to a (good) line-of-sight situation. Results for measurements including also OLOS conditions can be found in \cite{SDCpaper}. 
The distance between the UE and BS is 28\,\si{m}, wherein the LOS distance may change by 4\,\si{cm} at different heights relative to the shortest distance. The path length of the ground reflections changes from 6.6\,\si{cm} to 34\,\si{cm} relative to the LOS distance. Hence, the difference between the LOS path and the ground reflection changes by almost 3 times the wavelength. Accordingly, constructive (in-phase) and destructive addition of the LOS and ground reflection is expected when the UE height changes.

As transmit signal, we selected a 5G compliant SRS signal with 100\,\si{MHz} bandwidth, the same signal as used for the simulation. The SRS was transmitted by the UE with an omni-directional antenna. 
The antenna of the receiving BS is characterized with a \ 60\textdegree beamwidth. 

For the evaluation, we derived the reference signal received power (RSRP), reference signal received path power for the estimated first path (RSRPP) and the measured time-of-arrival (ToA) for the first path. 
In addition, we recorded the bandwidth limited CIR corresponding to the correlation between the transmitted SRS and the reference signal at the receiver.

Due to the low delay of the ground reflection it is difficult to distinguish between the LOS path and the ground reflection. Taking into account the in-phase and out-of-phase addition of the two components, from the level variation of the first correlation peak the strength of the ECs, in the experiment mainly the GR, can be estimated. 

\begin{figure}[t]
    \centering
    \includegraphics[width=1\columnwidth] {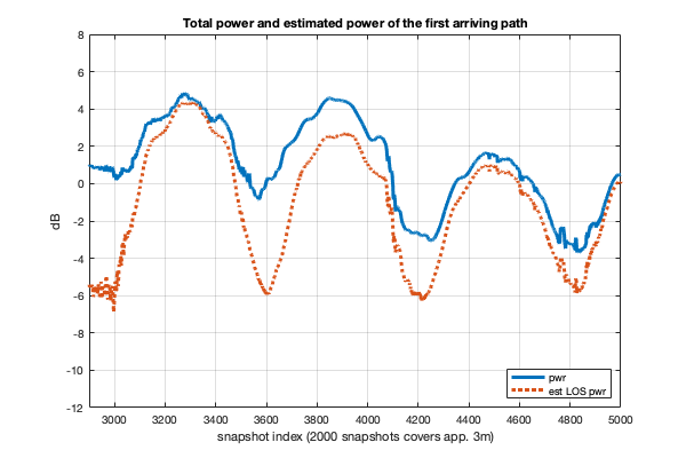}
    \caption{Plots of the measured total power and estimated LOS power vs.\ time from the measurement setup.}
    \label{fig:MeasurementPwr}
\end{figure}

Fig.~\ref{fig:MeasurementPwr} depicts the measured (total) signal power and estimated power of the first arriving path over a height variation of around 3\,\si{m} between the first and last measured snapshot. 
It can be observed that for the total power and estimated LOS power, the change of the phase difference between the direct signal and the reflected signal causes a high signal level variation. 

\begin{figure}[htbp]
    \centering
    \includegraphics[width=1\columnwidth] {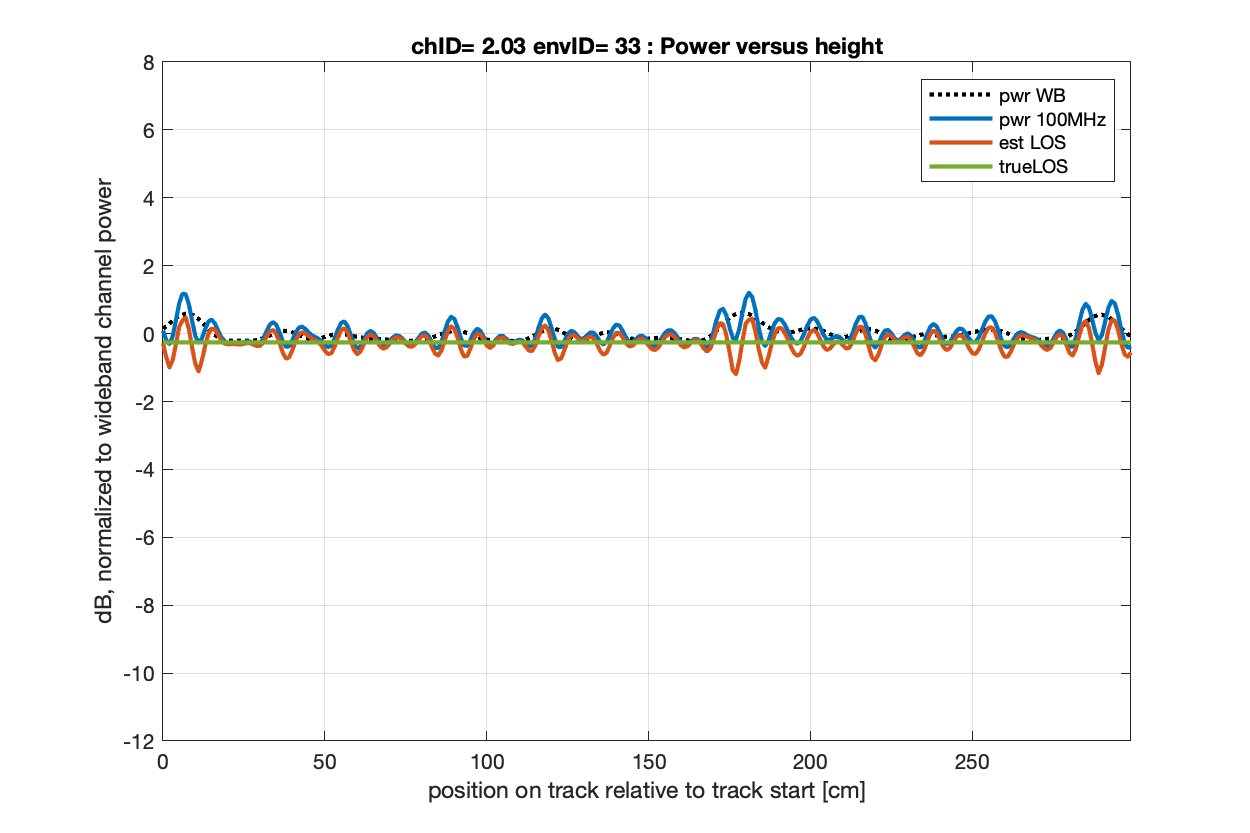}
    \caption{Plots of the measured total power and estimated LOS power vs.\ time from the InF-LOS simulation  setup without GR according to \cite{tr38901}.}
    \label{fig:SimSoAPwr}
\end{figure}

\begin{figure}[htbp]
    \centering
    \includegraphics[width=1\columnwidth] {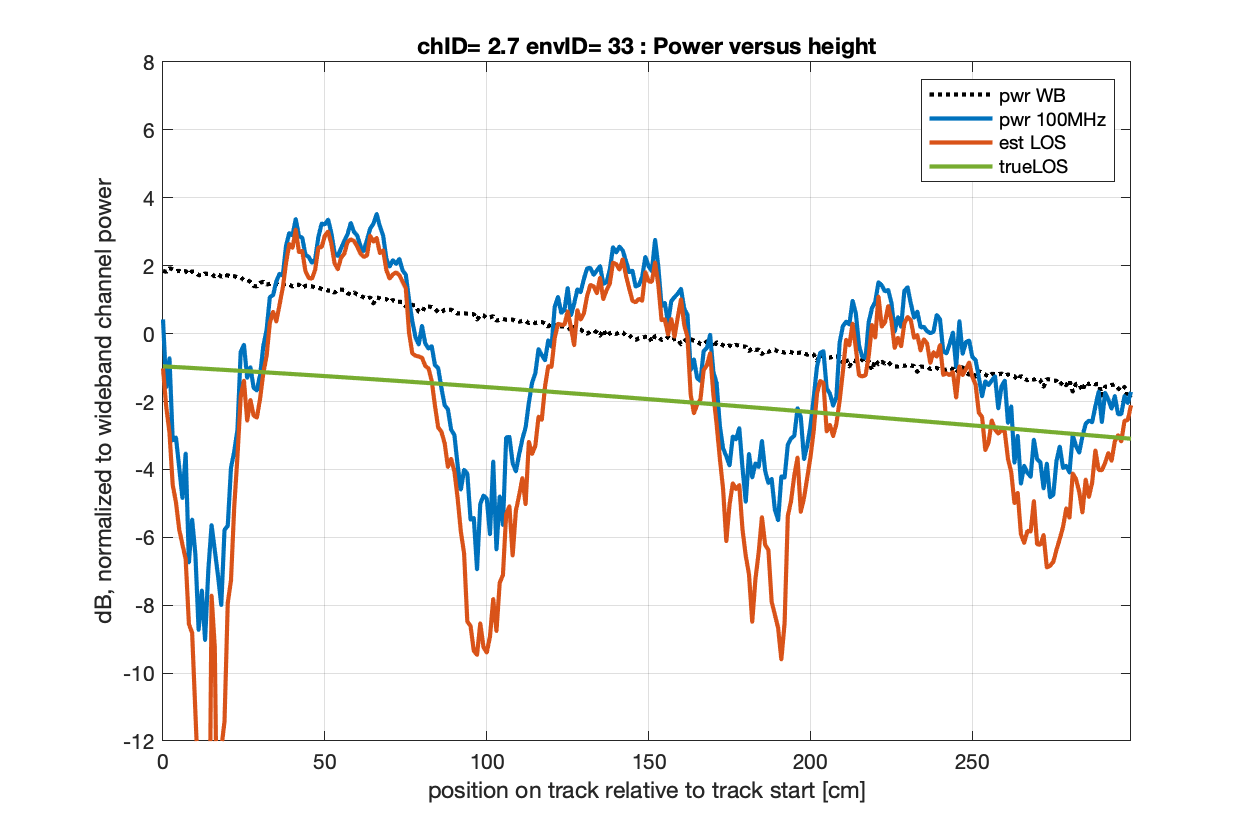}
    \caption{Plots of the measured total power and estimated LOS power vs.\ time from the InF-LOS simulation setup with EC (mainly GR).}
    \label{fig:SimPwr}
\end{figure}

We rebuild the setup using the model and compared the CIRs generated by the model with the measurements.
Fig.~\ref{fig:SimSoAPwr} shows the behavior using the model as defined by \cite{tr38901}. The optional feature ground reflection was NOT enabled. The spatial consistency procedure as defined by \cite{tr38901} was enabled for this simulation. 
It is observed that the model predicts a nearly constant signal power level. This is caused by the normalisation of the sum of the cluster power according the pathloss and the randomly generated shadow fading. The shadow fading is generated by the large scale model and is spatially correlated. Hence, for different UE heights the shadow fading is constant resulting in constant overall power. The observed small power variation result from the measurement of the power in the signal bandwidth, whereas the sum of the cluster power covers the ``wideband'' signal power. 
Fig.~\ref{fig:SimPwr} depicts the results for the simulation with EC (mainly GR in the example). The low delay of the EC cause a constructive and destructive addition of the LOS and EC within the bandwidth of 100\,\si{MHz}. Already without fine-tuning of the parameters a good match with the measurement is achieved.

\begin{figure*}[ht]
    \centering
    \subfigure {\includegraphics[width=1\columnwidth]{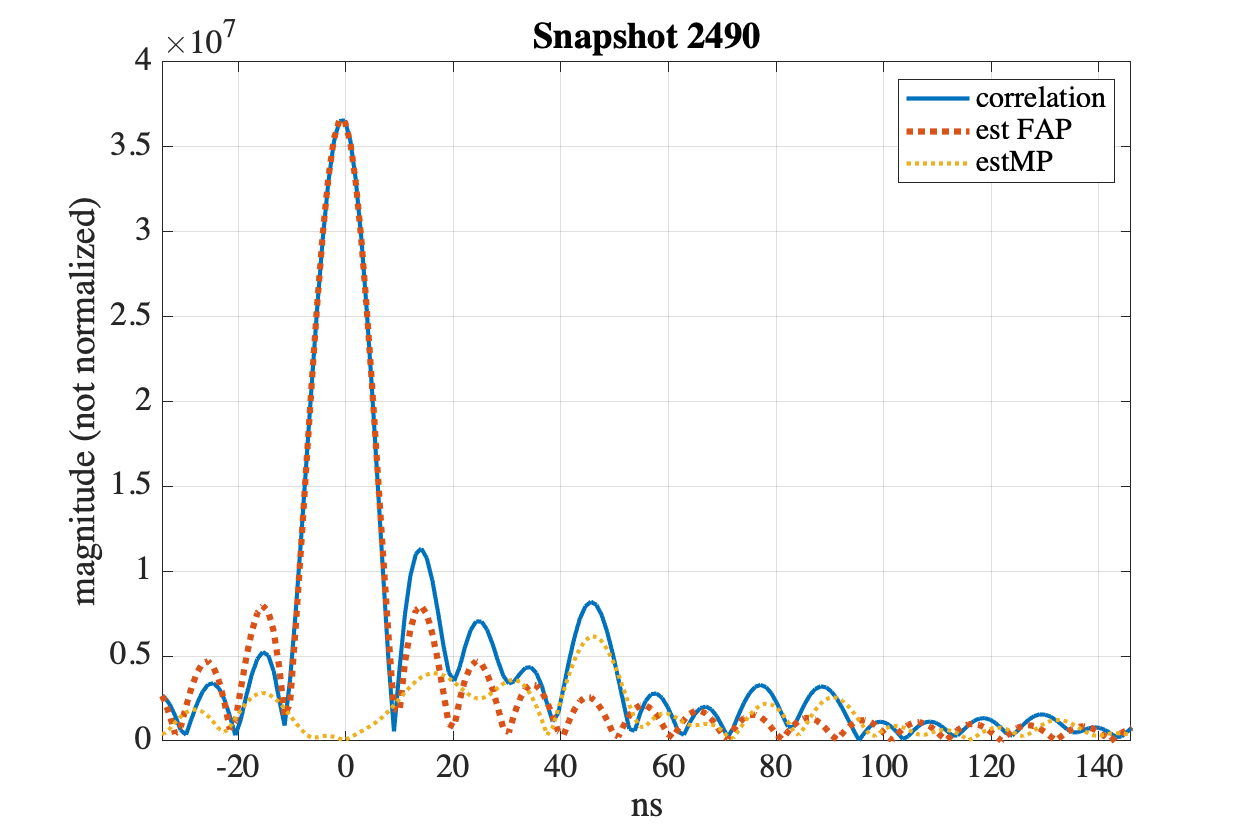}}
    \subfigure {\includegraphics[width=1\columnwidth]{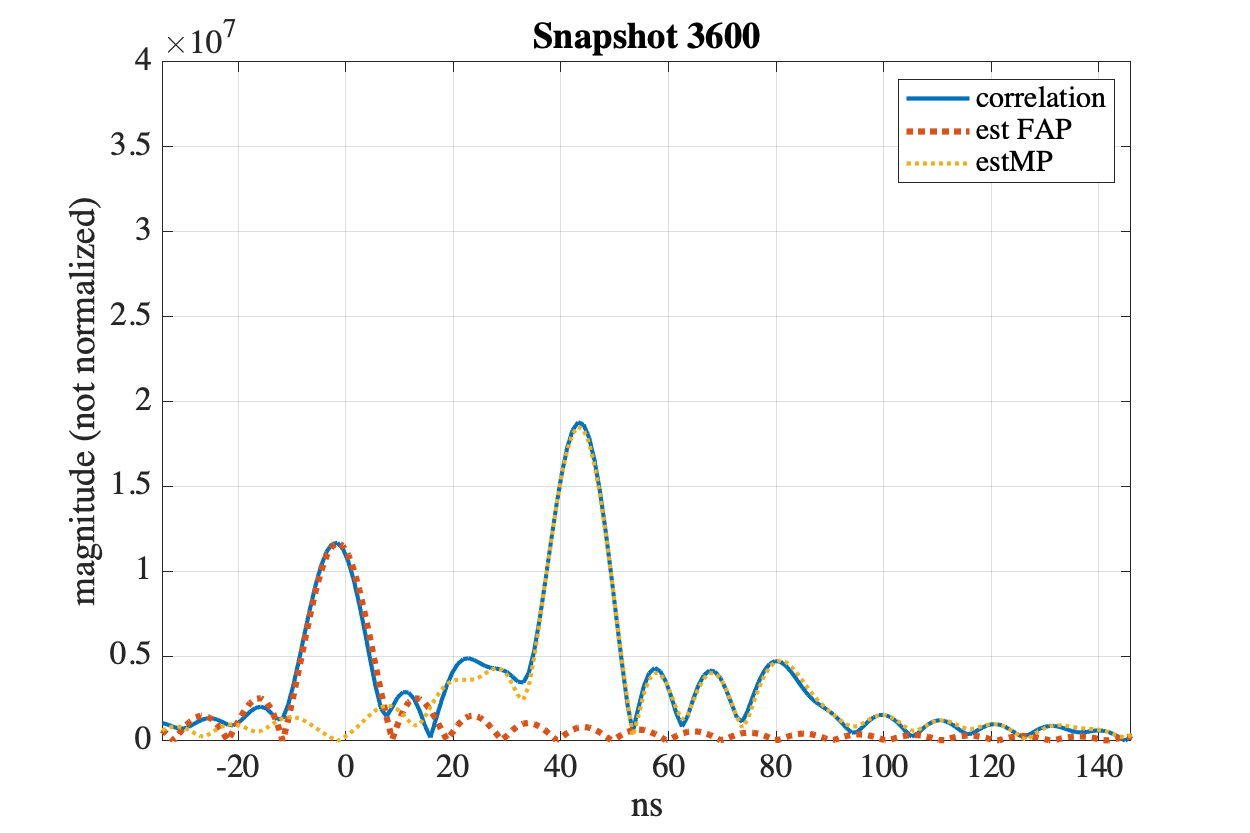}}
    \caption{Observed constructive (left) and destructive (right) contribution of the GR to LOS.}
    \label{fig:measConDes}
\end{figure*}

\begin{figure*}[ht]
    \centering
    \subfigure {\includegraphics[width=1\columnwidth]{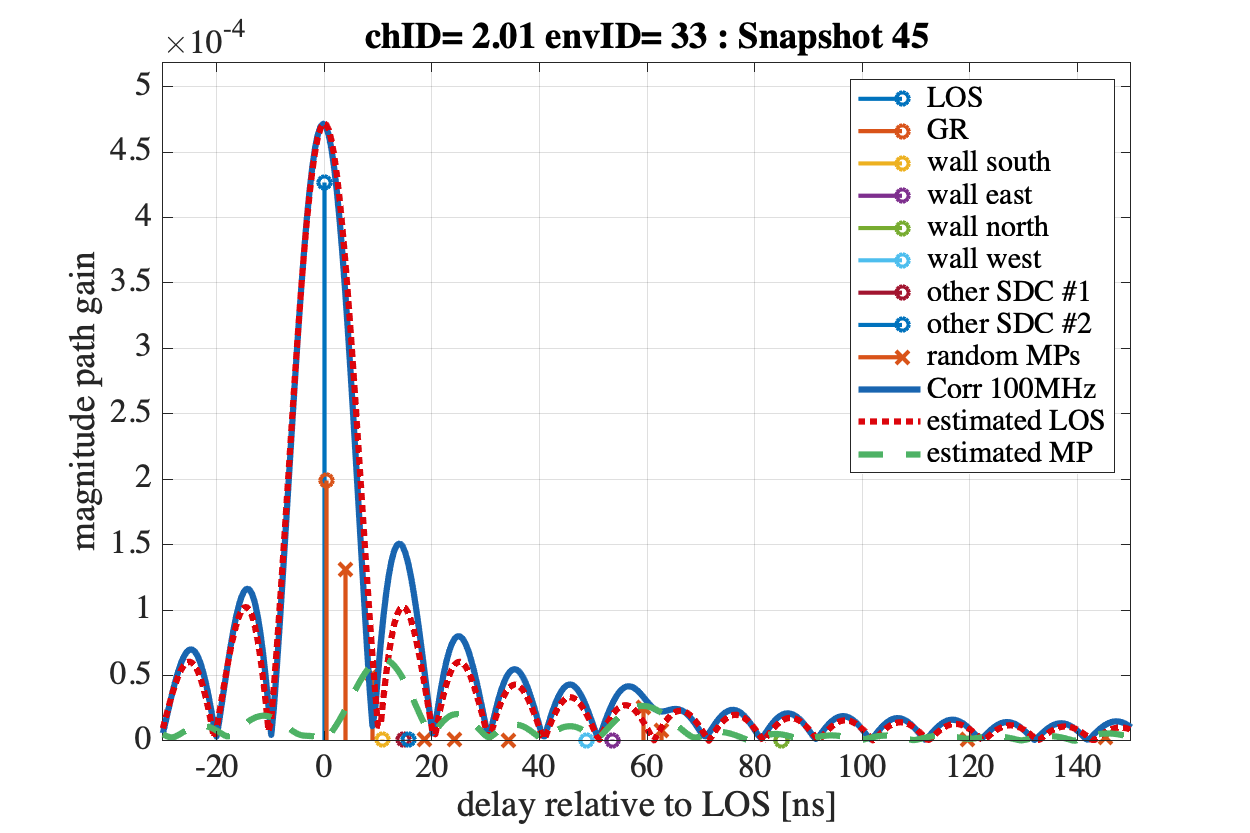}}
    \subfigure {\includegraphics[width=1\columnwidth]{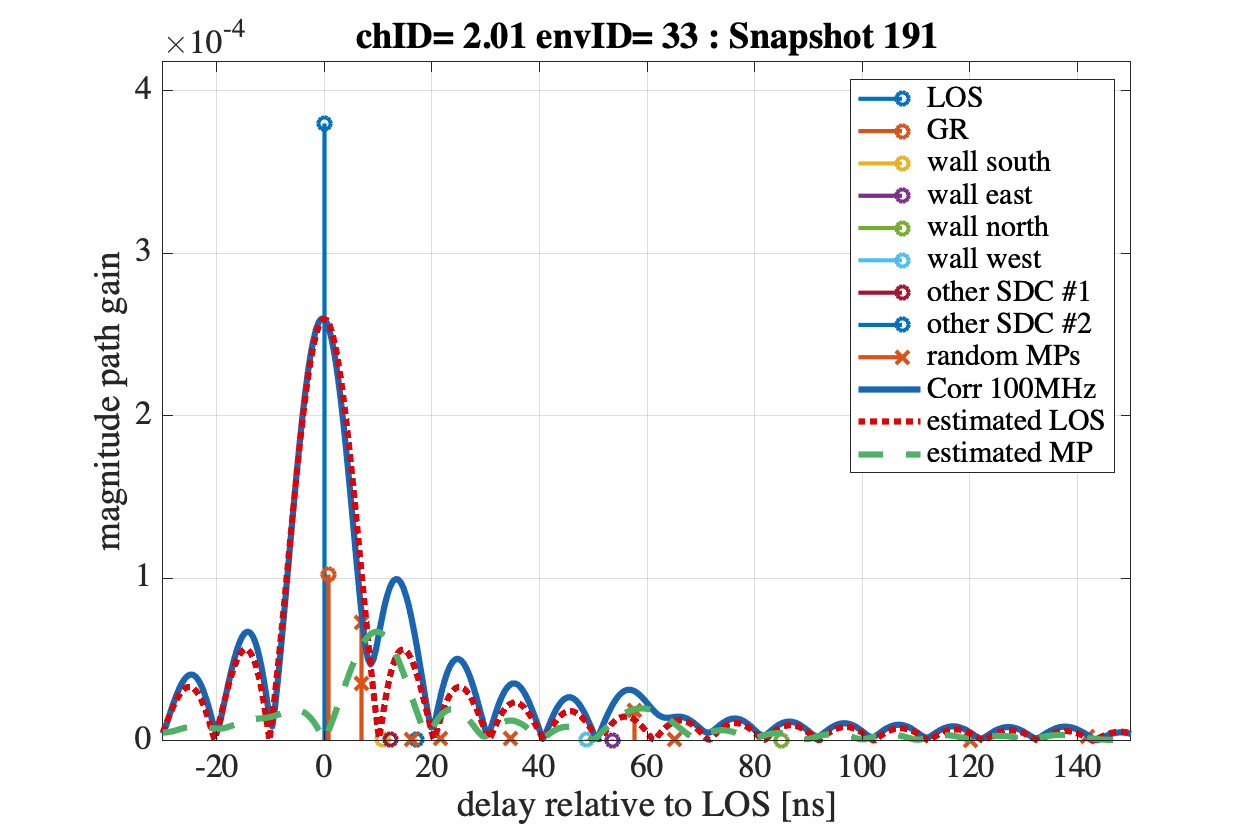}}
    \caption{Simulated Setup showing constructive (left) and destructive (right) contribution of the GR to LOS.}
    \label{fig:SDCConDes}
\end{figure*}

Recording the correlation function allows also a detailed comparison of the CIR generated in simulation and measured. As an example Fig.~\ref{fig:measConDes} depicts two examples of the measured CIRs and demonstrates the results from the constructive (in-phase) and destructive (opposite phase) superposition according to the strength of the GR on correlation function given the applied bandwidth-limited SRS. Due to the low propagation delay difference between the LOS and GR signal (less than 1\,\si{ns}) and the bandwidth of 100\,\si{MHz}, it is difficult to separate the signals. Hence, the measured first-arriving-path (FAP) power is the sum of the LOS and GR components. 
For the same positions we generated the CIRs also in simulation. In the simulation we know the power and phase of each cluster and can compare the true (wideband) CIR (WB-CIR) with the correlation function representing the bandwidth limited CIR. Fig.~\ref{fig:SDCConDes} shows the magnitude of the WB-CIR and the correlation associated to measured examples given in Fig.~\ref{fig:measConDes}. The figure show constructive addition of the LOS and GR (the magnitude of the correlation function is higher than the LOS component of the WB-CIR) and destructive superimposition (the correlation peak is lower than the magnitude of the LOS component of the WB-CIR). Comparing the simulation and the measured correlation we observe a significant difference at a delay close to 45\,\si{ns}. In the measurement this is caused from the reflection at the wall behind the UE. To include also these effects in the model a "hybrid approach" is required generating the position of reflecting cluster in a deterministic way. This additional extension of the TR38.901 model is subject of \cite{SDCpaper}  

 % \section{Characteristics of 38.901 model}
 % this is covered by the sections above

%%%%%%%%%%%%%%%%%% conclusion %%%%%%%%%%%%%%%%%%%%%%%
\section{Conclusion}
\label{sec:conclusion}
\noindent
In this paper, the effect of early clusters (EC) on radio channel modeling for simulations is presented. A precise modeling of the EC is essential for the verification and optimisation of algorithms targeting reality-like performance. Using channel model described in TR 38.901 as a baseline, a statistical analysis was performed. 
Due to the stochastic nature of the generation of the clusters and their respective parameters, we show that clusters representing strong reflections arriving with a low delay are generated with a low probability only. Hence, simulation results based on this model may be optimistic and may not justify the need of enhancements.

Complementary to the proposed model extension, a method for the statistical evaluation of the simulation results is presented. The proposed metric $K_{\mathsf{EC}}$ is a good indicator on the level of difficulty of the randomly generated CIRs and can be used for sorting the simulation results to better identify critical scenarios. 

Accordingly, for positioning evaluation based on TR38.901 we recommend the following:
\begin{itemize}
    \item Activation of optional features like ground reflection as already covered by the TR 38.901 model 
    \item More flexibility in the definition of the statistical properties of the power delay profile (PDP) is required. The proposed method combining two channels, whereas the first channel focuses on the parameters relevant for communication and the second channel covers the effects important for positioning, offers sufficient flexibility and allows the reuse of the existing channel model software with minor modifications.
   \item Further extensions are proposed in \cite{SDCpaper}.  
\end{itemize}

%\newpage\noindent % page balancing in paragraph%
% \newpage % page balancing between paragraphs

%For testing of the reliability of a method the channel model shall generate a sufficient number of links to reflect the different situation. Especially effects like ground reflection and non-ideal antenna pattern have a high impact to the effective KFEC. In industrial like scenarios many reflecting objects may be placed near to the UE, the probability of KFEC should be represented in the model. 

%Methods using the signal power as an indicator may work in simulation setups using 38.901 models without extension, but may fail for real-world scenario. With EC modelling a better match with real-world data can be achieved. 
%- "full statistical" and "deployment EXAMPLE" based setups with ray tracing have pros and cons
%- method adds advantages of ray tracing based method to statistical model 
%- Advantages (randomized CIRs) of statistical models are maintained   

%- core elements of 38.901 are maintained 
%- process relationship  AoA/AoD/pathdely  <=> cluster position is inverted 

%- statistical model for SDC position is FFS
% discuss and name outlook, what other aspects are missing and need to be addressed for "positioning channel models"%

%%%%%%%%%%%%%%%%%% references %%%%%%%%%%%%%%%%%%%%%%%
\bibliographystyle{ieeetr}%
\bibliography{bibliography}%

\end{document}